# Enhanced photocatalytic activation of methane C-H bond by Incorporate Cerium into ETS-10

*Cao Yang*

Department of Energy and Resources Engineering, College of Engineering, Peking University, Beijing, 100871 P. R. China

Email: caoyangchem@pku.edu.cn

ABSTRACT Ce incorporated ETS-10 was prepared and characterized, and the Cerium incorporation induce photocatalytic coupling of methane has been studied in detail. Ce modification enhanced the photocatalytic methane conversion performance, and Ce incorporation change the structure of ETS-10 significantly, especially Si chemical environment, after examined with various characterization. The SPS (surface photovoltage spectra) results show that the Cerium modification induce photovoltage, which means the strong electron-hole separation which enhance the photocatalytic performance. The charge-hole separation is non-traditional n-n heterojunction facilitate the charge-hole separation is highly oriented and efficient at atomic scale, the linear -Ti-O-Ti- and radical distribute Ce species form a n-n type heterojunction structure. This mechanism shows significant enhancement for methane conversion reaction.

MANUSCRIPT TEXT

Introduction

$CH_4$, just like coal or petroleum[1], as a widely distributed raw carbon resource, has always been regarded as a promising stocks to upgrade to versatile valuable chemicals. Transform methane into more complexed hydrocarbons without any extra oxidant were not easy task because of the stubborn C-H bond in methane. Traditional methane conversion mostly performed at high temperature, which consume large amount of energy, facile to get coked, catalyst lose activity, which is caused by over-reaction of methane at high temperature. Fortunately, photocatalysis could be another choice, when catalyst were caught by light, renewable hole oxidant would be formed, when methane react with hole oxidant, there is no enough heat to pass over energy barrier to generate coke when perform the reaction at room temperature.[2-3]

Cerium is a rare earth has excellent catalytic performance and have been applied in various fields. $CeO_2$ compound has exhibit photocatalysis performance on $CO_2$ reduction[4], $N_2$ fixation[5], water splitting[6]. Ce has been applied in anode materials in methane fuel cell[7]. It has been applied in light alkane oxidation and dehydration reaction[8]. $CeO_2$ doped with noble metal like Pd, Pt, Rh have been studied in alkane oxidation, dehydrogenation[9-11]. Front-row transition metals like Ni were also doped on $CeO_2$ to drive methane conversion reaction[12]. Recently, homogeneous $Ce^{3+}$ has exhibit photocatalytic performance on conversion of methane[13]. ETS-10 is a titanium-containing molecular sieve for its unique 1D atom Ti-O-Ti linkage structure with quantum confinement effect[14] and its can be written as $Na_nK_{2-n}TiSi_5O_{13}$[15]. $Ga_2O_3$ and Ga-ETS-10 has been developed in photocatalysis coupling of methane[16-17]. Ce and Ti have been modified on $SiO_2$'s surface to drive non-oxidative methane conversion and the author identified Ce will drive photocatalytic methane conversion, but the conversion performance (~0.127 % $CH_4$ conversion ratio) is far from satisfactory[18]. However, the photocatalysis ability of cerium incorporated in ETS-10 has not been reported yet. The relative study exhibits the hetero metal ion modification induce significant changes in micro-structure of ETS-10[14, 19]. The structural and optical properties were characterized and details of the photocatalytic performance were studies carefully. The plausible mechanism was put forward. The catalytic performance shows that the Cerium improves the catalytic methane conversion performance, it can be inferred that the elevated catalytic performance comes from efficient electron-hole separation induced by Ce-O-Ti's LMCT effect[20], the Ti relate quantum confinement effect of ETS-10 micropore structure[14, 21]. Compared with previous Ce/Ti doped $SiO_2$ catalyst, the methane conversion performance is magnitude elevated.

## Experimental Section

ETS-10 was prepared by reference, and some changes were made to improve the crystallinity and yield. The as-prepared ETS-10 was dispersed in $Ce(NO_3)_3$ solution with different concentration and stirred at room temperature overnight. Various lanthanide were incorporated into ETS-10. The detailed experimental results are depicted in the supporting information.

## Characterization

X-ray powder diffraction patterns were obtained with a JEOL RINT2000 vertical goniometer using Cu Kα radiation (λ=1.5418 Å). FT-IR spectroscopy were recorded with Perkin-Elmer GX spectrophotometer in the range 400 ~ 4000 $cm^{-1}$ with a resolution of 4 $cm^{-1}$, sample were grinded with dry KBr with a weight ratio around 1:5 and pressed under 10~20 MPa to form a compact transparent wafer. UV-Visible light absorption spectra were collected on a Hitachi U-4100 UV/Visible/Near Infrared Spectrophotometer (Solid). X-Ray Photoelectronic Spectroscopy was proceeded on the ThermoScientific ESCALAB250 with a $Ar^+$ gun. The catalytic performance data of $CH_4$ coupling were collected on a Shimadzu 14C gas chromatograph equipped with a FID detector. The column temperature was kept at 80 °C, the injector room temperature was kept at 160 °C, FID detector was set as 250 °C. The carrier gas was applied with high purity $N_2$.

# Results and discussion

The Raman spectroscopy of fresh prepared ETS-10 represent the well crystallinity of ETS-10[22]. And the XRD pattern clearly show that when cerium incorporation did not damage the origin framework structure of the ETS-10. Whereas the FTIR spectroscopy of Ce-ETS-10 and pristine ETS-10 show that the ~730 cm$^{-1}$ vibration peak descended significantly, which can be ascribed to the band stretching of Ti-O and 750 cm$^{-1}$ vibration peak is ascribed to the vibration band of ····-Ti-O-Ti-···· [14]. It can be inferred that Cerium incorporation induce significant changes towards the -Ti-O-Ti- environment.

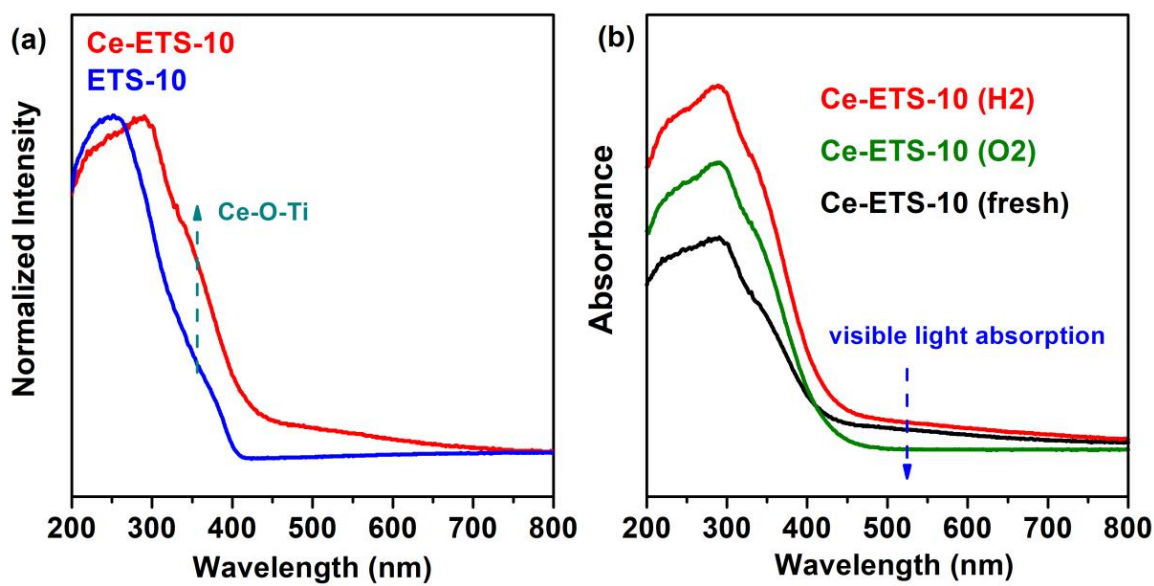

**Fig 1.** UV-Vis spectroscopy of (A) ETS-10, Ce-ETS-10, (B) pristine Ce-ETS-10 and Ce-ETS-10 treated in H$_2$, O$_2$.

**XRD spectroscopy** shows that pristine ETS-10 molecular sieve was well-crystallized without any impurities. When Cerium were incorporated into ETS-10's micropore, the structure was not damaged and the crystallites were hold as it was.

**The solid-state UV-Vis spectra** is changed when rare earth incorporated into ETS-10. Ce-ETS-10

exhibit absorption in the visible light region and show enhanced light absorption ability in the ultraviolet region between 350-400 nm$^{-1}$. That can be ascribed to be the light absorption by interaction between $TiO_3^{2-}$ and Cerium[23]. As a contrast, La, Y-ETS-10 does not show significant elevating adsorption in 350~400 nm, as we can infer that the strong electronic interaction between Ce and the ···Ti-O-Ti··· framework. And what's more, Ce-ETS-10 show a visible light absorption tail.

We treat Ce-ETS-10 by calcite in $H_2$ or $O_2$ atmosphere to make a better understanding of the interaction between Ce and ETS-10. When fresh Ce-ETS-10 is treated in hydrogen or photocatalysis with methane, it retains its original absorption in the visible light region. Whereas when treated in the oxygen atmosphere at about 250 °C, visible light absorption tail disappeared, it is considered that Cerium has affinity with oxygen and deduced the strong interaction between cerium and the framework of ETS-10. And the XRD results show that calcinate in $H_2$ or $O_2$ did not damage the framework structure of ETS-10.

But it can be recognized by $^{29}$Si NMR measurement, since that chemical shift of Si are significantly shifted to higher region originate from the interaction between highly basic O atoms connect with Ti atom and rare earth ion. And at the same time, the structure of ETS-10 did not damaged.

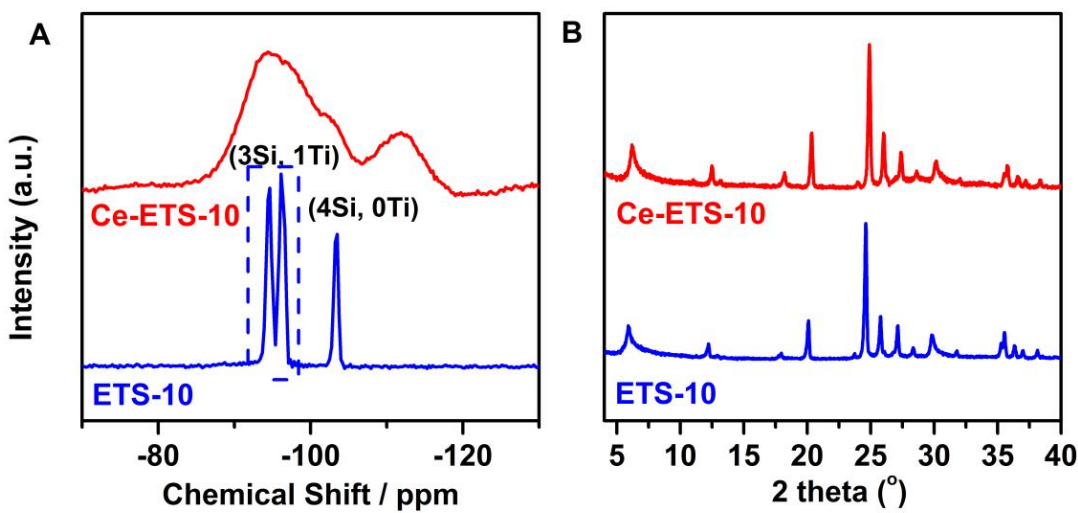

**Fig 2.** (A) $^{29}$Si NMR spectroscopy (B) XRD pattern of ETS-10, Ce-ETS-10.

**Solid state NMR**

$^{29}$Si MAS NMR were obtained on a Varian 400 MHz and all $^{29}$Si chemical shifts were referred to TMS via a Kaolin secondary standard. The data collected is consistent to that reported well-crystalline (Na, K)-ETS-10. The Figure shows that a group of four peaks between -94 and -97 ppm which due to silicon (3Si,1Ti) and another peak at -103.4 ppm due to silicon(4Si, 0Ti)[22]. When (Na, K)-ETS-10 is modified with Cerium, the peaks all moved towards higher chemical shift and the peak were broadened significantly, which can be explained by intense interaction between Cerium and Silicon compared with that interaction between Na$^+$ (or K$^+$) and Silicon in the framework of ETS-10，and this phenomenon is consistent with earlier report. Since that part of Na$^+$ and K$^+$ are substituted by Cerium, which have greater Electrostatic interaction with Si atom in the framework of ETS-10, subsequently cause higher chemical shift.

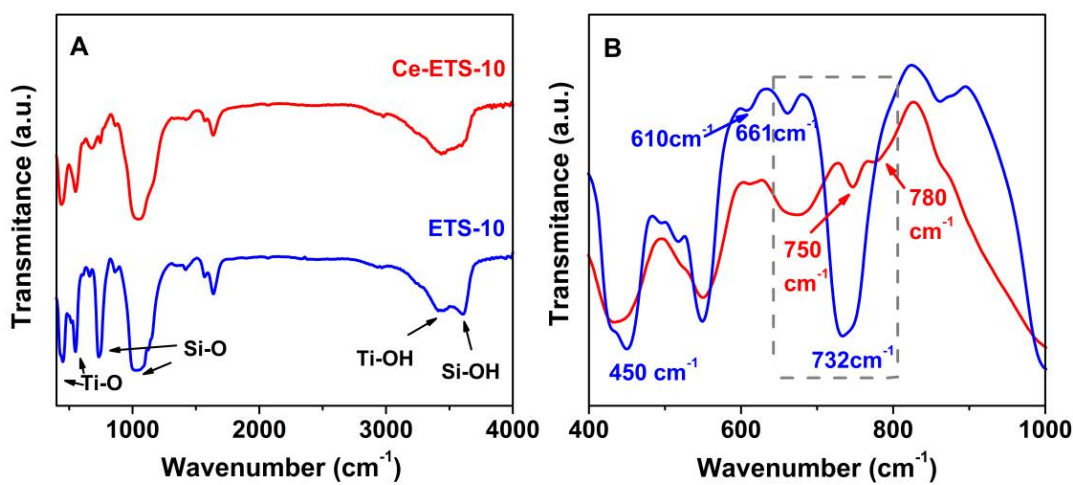

**Fig 3.** FT-IR spectroscopy of ETS-10, Ce-ETS-10 in (A) 4000~400 cm$^{-1}$ (B) 1000~400 cm$^{-1}$.

Significant difference can be tracked between fresh (Na, K)-ETS-10 and Cerium modified ETS-10, especially in 400 ~ 1000cm$^{-1}$. The IR absorption around 732 cm$^{-1}$ descend greatly and shifted about 18

cm$^{-1}$ to the higher frequency. The absorption at around 732 cm$^{-1}$ was ascribed to the stretching band of Ti-O, when fresh (Na, K)-ETS-10 is modified with Cerium, the bond vibration changes. As is has been already known that around 750 cm$^{-1}$ and 661 cm$^{-1}$ are always been ascribed to the vibration of …-Ti-O-Ti-… atom wire and vertical -Ti-O-Si- stretching vibration [14], the correlate IR intensity changed significantly, the vertical -Ti-O-Si- stretching vibration is stronger than …-Ti-O-Ti-… when cerium is introduced, and the …-Ti-O-Ti-… atoms wire stretching vibration intensity is depressed. The results clearly expressed cerium induced correlate interaction with …-Ti-O-Ti-… subunit.

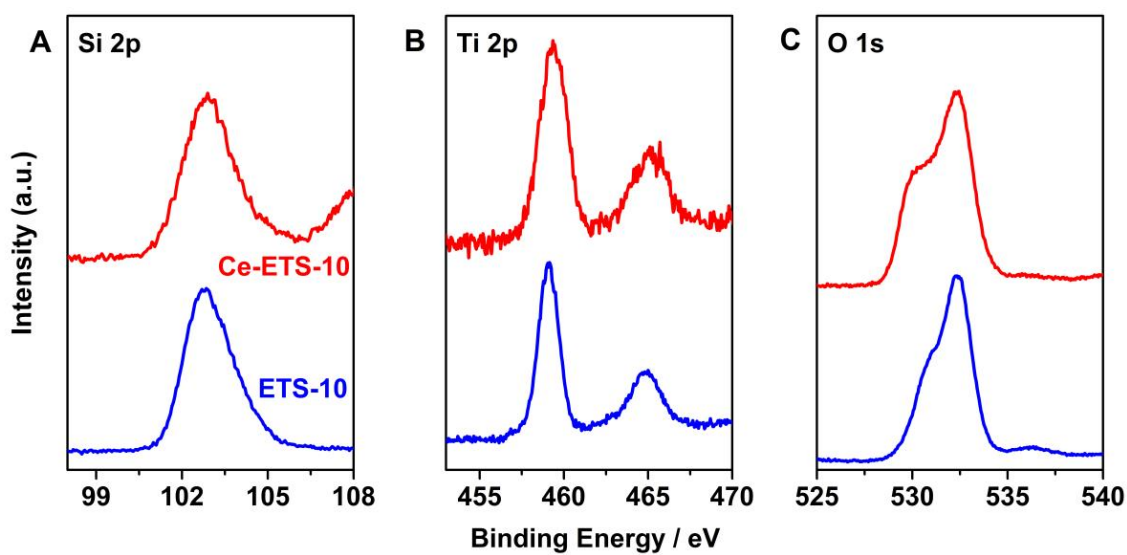

**Fig 4.** Si 2p, Ti 2p, O 1s XPS spectroscopy of ETS-10, Ce-ETS-10.

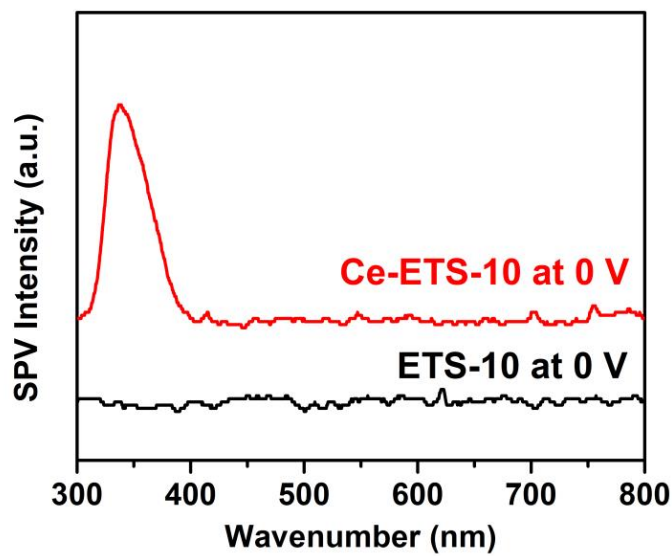

**Fig 5.** Surface photovoltage spectra of ETS-10, Ce-ETS-10.

**The SPS spectroscopy** also give another effective proof that ETS-10 incorporated with cerium to convenient the electron and holes separation even in the visible light region. It can be realized that there is photo voltage signal that extend to the visible light region which means that Ce-ETS-10 can efficiently separate electrons and holes even in the visible light region. Whereas pristine ETS-10 without cerium incorporated in did not show such photo voltage signal in the visible region. These characterizations make it clear that how Ce-ETS-10 exhibit photocatalysis ability in non-oxidization coupling of methane. And it can be recognized by us that the photocatalytic structure of Ce-O-Ti is easily established by just stirred in the room temperature compared to the complex procedures in the literature[23].

The valance of cerium in different amount of cerium incorporated in the framework of ETS-10 were also examined by **XPS spectroscopy**. It is an unintelligible phenomenon to understand that $Ce^{3+}$ can be found in the light incorporated ETS-10 whereas $Ce^{4+}$ is the more stable state and the main existence form when larger amount of cerium is incorporated in ETS-10 without further condition changed.

**Catalytic performance**

It can be seen from the catalytic data that Cerium incorporated ETS-10 have an extraordinary performance than that of La, Gd, Lu or Y incorporated ETS-10. It was supposed that the interaction between cerium and ETS-10 play an important role. Electron transform can be detected by SPS and ESR to make it clear that $Ce^{4+}$ can receive an electron and subsequently reduced to $Ce^{3+}$ when $TiO_3^{2-}$ was excited and electrons and holes are separated. The catalytic performance shows that Ce incorporation induce methane conversion performance, the Ce-ETS-10-2.5, Ce-ETS-10-100, Ce-ETS-10-200, Ce-ETS-10-250 sample exhibit methane conversion ratio are 1.124 %, 2.049 %, 3.173 %, 5.499 %, respectively. Ce-ETS-10-250 shows the best methane conversion performance and the corresponding methane conversion amount is 10.998 μmol, and the relate conversion ratio is around 9.17 μmol $g^{-1}$ $hr^{-1}$. La-ETS-10-200 and Y-EST-10-200 shows a methane conversion ratio at 0.557 % and 0.3042 % during the similar reaction process, respectively. Other rare earth metals like Pr, Nb, Pm, Sm, Gd, Sm, Lu modification shows nearly mere or no photocatalytic performance. We expect that the strong interaction between Cerium and framework of ETS-10 is responsible for the catalytic performance.

Cautions: The catalytic performance is sensitive to the structure of localized incorporated Cerium species, and the pristine ETS-10 should be carefully synthesized, the impurities in ETS-10 may influence Cerium incorporation. The detailed structure of Cerium incorporated in ETS-10 would be study in detail in the future, it shows that the incorporation condition will hold great effect on the catalytic performance.

Reference

1.	Schwach, P.; Pan, X.; Bao, X., Direct Conversion of Methane to Value-Added Chemicals over Heterogeneous Catalysts: Challenges and Prospects. *Chemical Reviews* **2017,** *117* (13), 8497-8520.
2.	Meng, X.; Cui, X.; Rajan, N. P.; Yu, L.; Deng, D.; Bao, X., Direct Methane Conversion under Mild Condition by Thermo-, Electro-, or Photocatalysis. *Chem* **2019,** *5* (9), 2296-2325.
3.	Yuliati, L.; Yoshida, H., Photocatalytic conversion of methane. *Chemical Society Reviews* **2008,**

*37* (8), 1592-1602.
4. Chang, K.; Zhang, H.; Cheng, M.-j.; Lu, Q., Application of Ceria in CO2 Conversion Catalysis. *ACS Catalysis* **2020,** *10* (1), 613-631.
5. Zhang, C.; Xu, Y.; Lv, C.; Bai, L.; Liao, J.; Zhai, Y.; Zhang, H.; Chen, G., Amorphous engineered cerium oxides photocatalyst for efficient nitrogen fixation. *Applied Catalysis B-Environmental* **2020,** *264*.
6. Dvořák, F.; Szabová, L.; Johánek, V.; Farnesi Camellone, M.; Stetsovych, V.; Vorokhta, M.; Tovt, A.; Skála, T.; Matolínová, I.; Tateyama, Y.; Mysliveček, J.; Fabris, S.; Matolín, V., Bulk Hydroxylation and Effective Water Splitting by Highly Reduced Cerium Oxide: The Role of O Vacancy Coordination. *ACS Catalysis* **2018,** *8* (5), 4354-4363.
7. Murray, E. P.; Tsai, T.; Barnett, S. A., A direct-methane fuel cell with a ceria-based anode. *Nature* **1999,** *400* (6745), 649-651.
8. Campbell, K. D.; Zhang, H.; Lunsford, J. H., Methane activation by the lanthanide oxides. *The Journal of Physical Chemistry* **1988,** *92* (3), 750-753.
9. Haneda, M.; Mizushima, T.; Kakuta, N., Synergistic Effect between Pd and Nonstoichiometric Cerium Oxide for Oxygen Activation in Methane Oxidation. *The Journal of Physical Chemistry B* **1998,** *102* (34), 6579-6587.
10. Xie, P.; Pu, T.; Nie, A.; Hwang, S.; Purdy, S. C.; Yu, W.; Su, D.; Miller, J. T.; Wang, C., Nanoceria-Supported Single-Atom Platinum Catalysts for Direct Methane Conversion. *ACS Catalysis* **2018,** *8* (5), 4044-4048.
11. Bai, S.; Liu, F.; Huang, B.; Li, F.; Lin, H.; Wu, T.; Sun, M.; Wu, J.; Shao, Q.; Xu, Y.; Huang, X., High-efficiency direct methane conversion to oxygenates on a cerium dioxide nanowires supported rhodium single-atom catalyst. *Nature Communications* **2020,** *11* (1), 954.
12. Akri, M.; Zhao, S.; Li, X.; Zang, K.; Lee, A. F.; Isaacs, M. A.; Xi, W.; Gangarajula, Y.; Luo, J.; Ren, Y.; Cui, Y.-T.; Li, L.; Su, Y.; Pan, X.; Wen, W.; Pan, Y.; Wilson, K.; Li, L.; Qiao, B.; Ishii, H.; Liao, Y.-F.; Wang, A.; Wang, X.; Zhang, T., Atomically dispersed nickel as coke-resistant active sites for methane dry reforming. *Nature Communications* **2019,** *10* (1), 5181.
13. Hu, A.; Guo, J.-J.; Pan, H.; Zuo, Z., Selective functionalization of methane, ethane, and higher alkanes by cerium photocatalysis. *Science* **2018,** *361*, eaat9750.
14. Jeong, N. C.; Lee, Y. J.; Park, J.-H.; Lim, H.; Shin, C.-H.; Cheong, H.; Yoon, K. B., New Insights into ETS-10 and Titanate Quantum Wire: A Comprehensive Characterization. *Journal of the American Chemical Society* **2009,** *131* (36), 13080-13092.
15. Anderson, M. W.; Terasaki, O.; Ohsuna, T.; Philippou, A.; MacKay, S. P.; Ferreira, A.; Rocha, J.; Lidin, S., Structure of the microporous titanosilicate ETS-10. *Nature* **1994,** *367* (6461), 347-351.
16. Li, L.; Cai, Y.-Y.; Li, G.-D.; Mu, X.-Y.; Wang, K.-X.; Chen, J.-S., Synergistic Effect on the Photoactivation of the Methane C H Bond over Ga3+-Modified ETS-10. *Angewandte Chemie International Edition* **2012,** *51* (19), 4702-4706.
17. Yuliati, L.; Hattori, T.; Itoh, H.; Yoshida, H., Photocatalytic nonoxidative coupling of methane on gallium oxide and silica-supported gallium oxide. *Journal of Catalysis* **2008,** *257* (2), 396-402.
18. Leny, Y.; Hideaki, I.; Hisao, Y., Modification of Highly Dispersed Cerium Oxides on Silica with Highly Dispersed Titanium Oxides as a New Photocatalyst Design for Nonoxidative Direct Methane Coupling. *Chemistry Letters* **2006,** *35* (8), 932-933.
19. Pavel, C. C.; Zibrowius, B.; Löffler, E.; Schmidt, W., On the influence of ion exchange on the local structure of the titanosilicate ETS-10. *Physical Chemistry Chemical Physics* **2007,** *9* (26), 3440-3446.
20. Chen, Y.; Du, J.; Zuo, Z., Selective C-C Bond Scission of Ketones via Visible-Light-Mediated Cerium Catalysis. *Chem* **2020,** *6* (1), 266-279.
21. Grommet, A. B.; Feller, M.; Klajn, R., Chemical reactivity under nanoconfinement. *Nature Nanotechnology* **2020,** *15* (4), 256-271.
22. Southon, P. D.; Howe, R. F., Spectroscopic Studies of Disorder in the Microporous Titanosilicate ETS-10. *Chemistry of Materials* **2002,** *14* (10), 4209-4218.

23. Nakamura, R.; Okamoto, A.; Osawa, H.; Irie, H.; Hashimoto, K., Design of All-Inorganic Molecular-Based Photocatalysts Sensitive to Visible Light: Ti(IV)−O−Ce(III) Bimetallic Assemblies on Mesoporous Silica. *Journal of the American Chemical Society* **2007,** *129* (31), 9596-9597.

**Electronic Supplementary Information (ESI):**

**Enhanced photocatalytic activation of methane C-H by Incorporate Cerium to ETS-10**

**Yang Cao**

[a] Department of Energy and Resources Engineering, College of Engineering, Peking University, Beijing, 100871 PR China

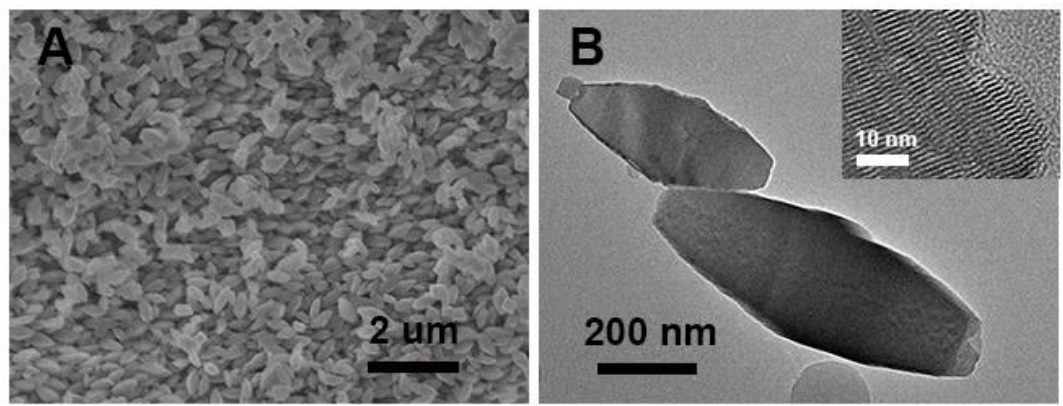

Figure S1. SEM and HRTEM spectra of ETS-10

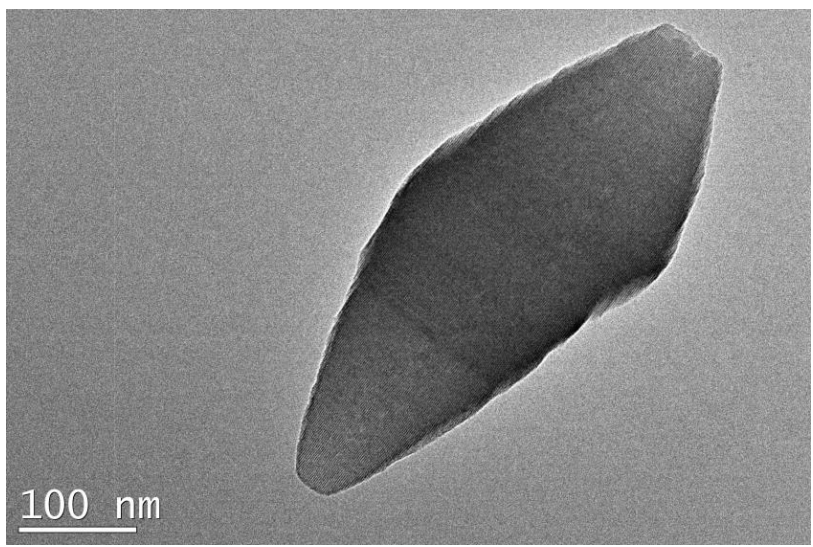

Figure S2. TEM spectra of ETS-10

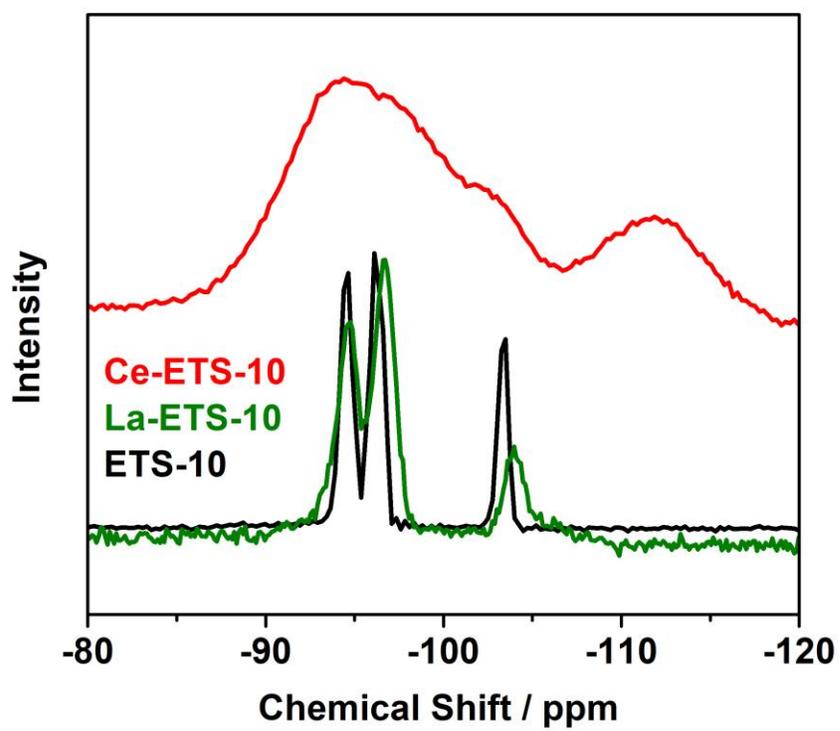

Figure S3. $^{29}$Si NMR spectra of ETS-10, La-ETS-10, Ce-ETS-10

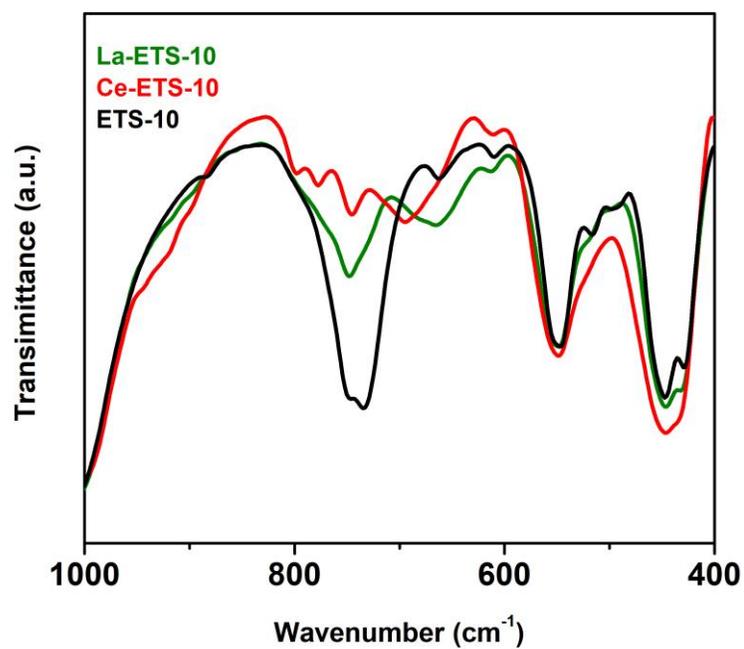

Figure S4. FT-IR spectra of ETS-10, La-ETS-10, Ce-ETS-10 in 400~1000 cm$^{-1}$.

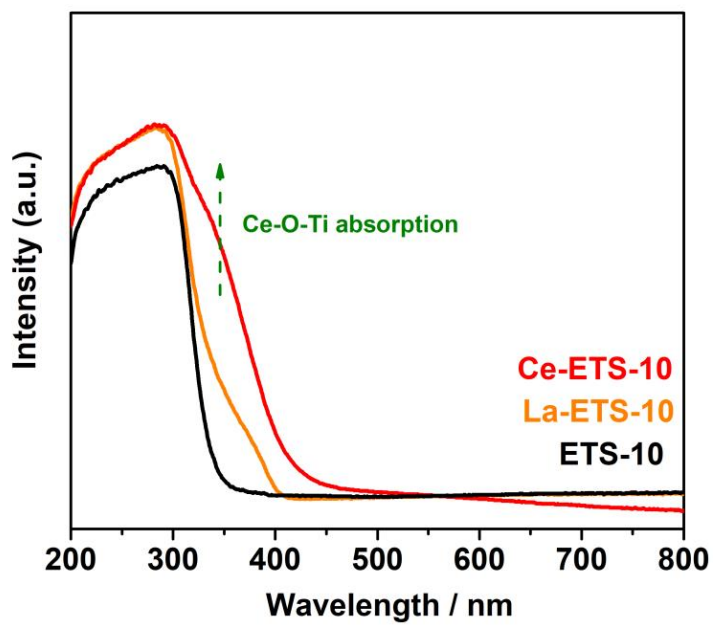

Figure S5. UV-Vis spectra of ETS-10, La-ETS-10, Ce-ETS-10. The arrow represents the Ce-O-Ti absorption.

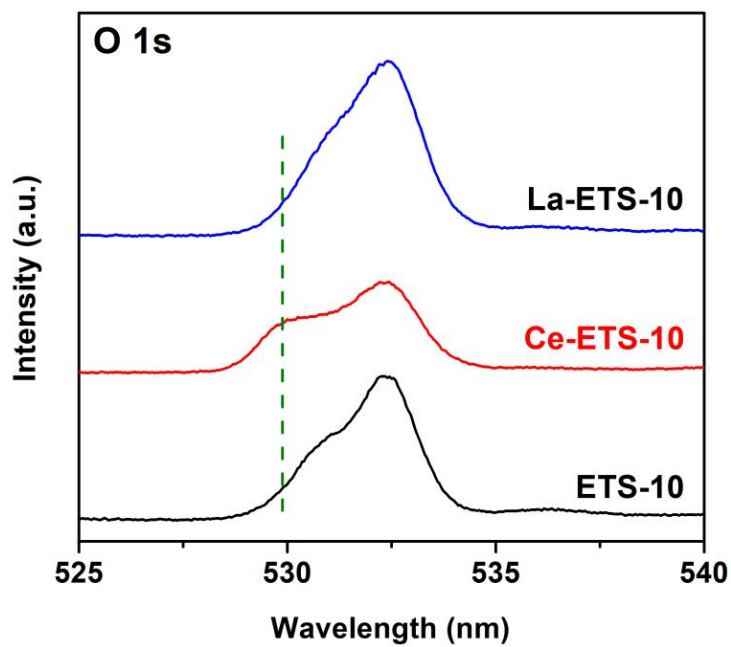

Figure S6. O 1s XPS spectra of ETS-10, La-ETS-10, Ce-ETS-10. The dashed line clarify the shoulder peak of Ce-ETS-10.

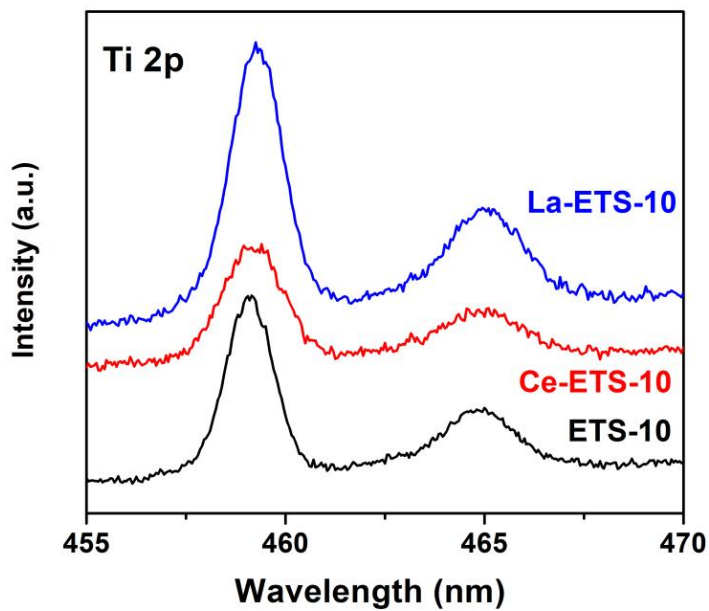

Figure S7. Ti 2p XPS spectra of ETS-10, La-ETS-10, Ce-ETS-10.

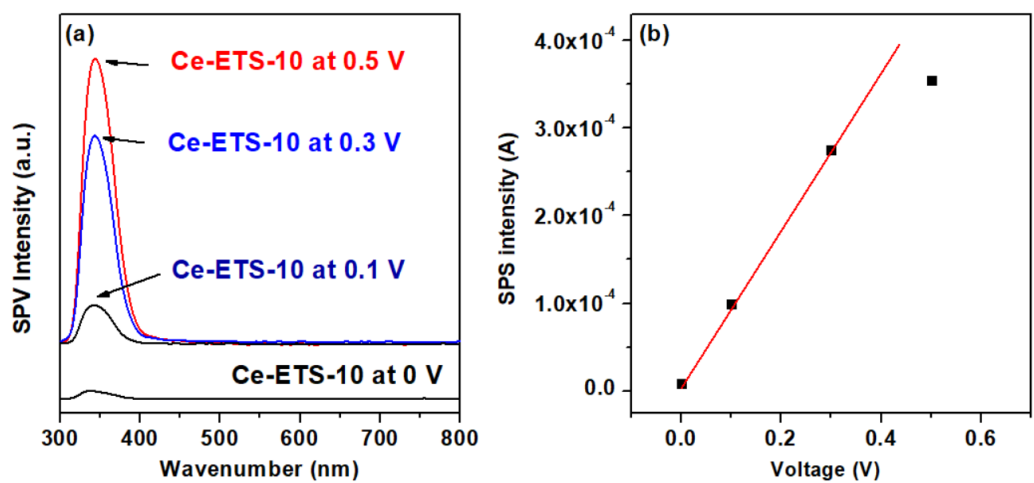

Figure S8. Surface photovoltage spectra of ETS-10, Ce-ETS-10 at biased voltage.

Table S1. FT-IR Vibration footprint spectra of ETS-10 (1500~400 cm$^{-1}$).

| Characteristic band | Characteristic band (cm$^{-1}$) |
| --- | --- |
| Si–O stretching | 1025 |
| O–Si–O bending | 435 |
| Ti–O stretching | 668, 729 |
| O–Ti–O bending | 435, 545 |